\documentclass[%
 reprint,
 amsmath,amssymb,
 aps,
]{revtex4-2}

\usepackage{comment} 
\usepackage{siunitx} 
\usepackage[utf8]{inputenc}
\usepackage{xparse} 
\usepackage{physics} 
\usepackage{graphicx}
\usepackage{dcolumn}
\usepackage{bm}
\usepackage{natbib,hyperref}

\begin{document}

\preprint{APS/123-QED}

\title{
Compact localised states in magnonic Lieb lattices
}

\author{Grzegorz Centała}
\author{Jarosław W. Kłos}%
 \email{klos@amu.edu.pl}
\affiliation{%
Institute of Spintronics and Quantum Information, Faculty of Physics, Adam Mickiewicz University, Poznań, Uniwersytetu Poznańskiego 2, Poznań 61-614, Poland
}%

\date{March 26, 2023} 

\begin{abstract}

Lieb lattice is one of the simplest bipartite lattices where compact localized states (CLS) are observed. This type of localisation is induced by the peculiar topology of the unit cell, where the modes are localized only on one sublattice due to the destructive interference of partial waves. The CLS exist in the absence of defects and are associated with the flat bands in the dispersion relation. The Lieb lattices were successfully implemented as optical lattices or photonic crystals. This work demonstrates the possibility of magnonic Lieb lattice realization where the flat bands and CLS can be observed in the planar structure of sub-micron in-plane sizes. Using forward volume configuration, we investigated numerically (using the finite element method) the Ga-dopped YIG layer with cylindrical inclusions (without Ga content) arranged in a Lieb lattice of the period 250~nm. We tailored the structure to observe, for the few lowest magnonic bands, the oscillatory and evanescent spin waves in inclusions and matrix, respectively. Such a design reproduces the Lieb lattice of nodes (inclusions) coupled to each other by the matrix with the CLS in flat bands.

\begin{description}
\item[Keywords]
flat bands, compact localized states, Lieb lattice, spin waves, finite element method
\end{description}
\end{abstract}

\maketitle


\section{\label{sec:level1} Introduction }
There are many mechanisms leading to wave localization in systems with long-range order, i.e. in crystals or quasicrystals. The most typical of these require (i) the local introduction of defects, including the defects in the form of surfaces or interfaces \cite{Davison_1996} (ii) the presence of global disorder \cite{Abrahams_2010}, (iii) the presence of external fields \cite{Wannier_1962} or (iv) the existence of many-body phenomena \cite{Abanin_2019}. However, since at least the late 1980s, it has been known that localization can occur in unperturbed periodic systems in the absence of fields and many-body effects, and is manifested by the presence of flat, i.e., dispersion-free bands in the dispersion relation. The pioneering works are often considered to be the publications of B. Sutherland \cite{Sutherland_1986} and E. H. Lieb \cite{Lieb_1989}, who found the {\em flat bands} of zero energy \cite{Ezawa_2020} for bipartite lattices with use of the tight-binding model Hamiltonians, where the hoppings occur only between sites of different sublattices. The simplest realization of this type of system is regarded as the Lieb lattice \cite{Lieb_1989,Leykam_2018}, where the nodes of one square sublattice, of coordination number $z=4$, connect to each other only via nodes with a coordination number $z=2$ from other two square sublattices (Fig.~\ref{Fig:structure1}). In the case of extended Lieb lattices \cite{Bhattacharya_2019, Zhang_2017}, the nodes of $z=2$  form chains: dimmers, trimmers, etc.(Fig.~\ref{Fig:structure2}).  An intuitive explanation for the presence of the flat bands is the internal isolation of excitations located in one of the sublattices. The cancelling of excitations at one sublattice is the result of forming destructive interference and local symmetry within the complex unit cell \cite{Flach_2014}. When only one of the sublattices is excited, the other sublattice does not mediate the coupling between neighbouring elementary cells, and the phase difference between the cells is irrelevant to the energy (or the frequency) of the eigenmode on the whole lattice - i.e. the Bloch function.  Modes of this type are therefore degenerated for different wave vector values in infinite lattices. We are dealing here with the localization on specific arrangements of structure elements, which are isolated from each other. Such kinds of modes are called {\em compact localized states} (CLS) \cite{Aoki_1996, Bergman08, Xia2018,Rhim_2019,Rhim_2021} and show a certain resistance to the introduction of defects \cite{Leykam_2017,Chalker2010}.
The flat band systems with CLS are the platform for the studies of Anderson localization \cite{Leykam2013}, and unusual properties of electric conductivity \cite{Hausle_2015}.
A similar localization is observed in the quasicrystals, where the arrangements of the elements composing the structure are replicated aperiodically and self-similarly throughout the system \cite{Kohmoto_1987, Mieszczak_2022} and the excitation can be localized on such patterns. The CLS in finite Lieb lattices have a form of loops (plaquettes) occupying the majority nodes ($z=2$). These states are linearly dependent and do
not form a complete basis for the flat band. Therefore occupancy gaps need to be filled (for infinite lattice) by states occupying only one sublattice of majority nodes, localizes at lines, called {\em noncontractible loop states} (NLS) \cite{Xia2018,Rhim_2019,TangS2020}.   

The topic of Lieb lattices and other periodic structures with compact localization and flat bands was renewed \cite{Leykam_2018} about 10 years ago when physical realizations of synthetic Lieb lattices began to be considered for electronic systems \cite{Drost_2017, Slot_2017}, optical lattices \cite{Shen_2010, Taie_2015}, superconducting systems \cite{Swain_2020,Xu_2021}, in phononics \cite{Ma_2021} and photonics \cite{Vicencio_2015,Xia2018}.
In a real system, where the interaction cannot be strictly limited to the nearest elements of the structure, the bands are not perfectly flat. Therefore, some authors use the extended definition of the flat band to consider the bands that are flat only along particular directions or in the proximity of high-symmetry Brillouin zone points \cite{Cracknell_1973}. In tight-binding models, this effect can be included by considering the hopping to at least next-nearest-neighbours \cite{Beugeling_2012,Jiang2019}. Similarly, the crossing of the flat band by Dirac cones can be transformed into anti-crossing and lead to opening gaps, separating the flat band from dispersive bands. This effect can in induced by the introduction of spin-orbit term to tight-binding Hamiltonian (manifested by the introduction of Peierls phase factor to the hopping) or by dimerization of the lattice (by alternative changes of hoppings or site energies) \cite{Beugeling_2012, Jiang2019, Beli2017, Ramachandran_2017, jiang_lieb-like_2019}. The later scenario can be easily observed in real systems where the position of rods/wells (mimicking the sites of Lieb lattice) and contrast between them can be easily altered \cite{Poli_2017}. Opening the narrow gap between flat band and dispersive bands for Lieb lattice is also fundamentally interesting because it leads to the appearance of so-called Landau-Zener Bloch oscillations \cite{Khomeriki_2016}. 

The isolated and perfectly flat bands for Lieb lattices are topologically trivial -- their Chern number is equal to zero \cite{Chen_2014}. For weakly dispersive (i.e. almost flat) bands the Chern numbers can be non-zero \cite{BERGHOLTZ}. 
 However, when the flat band is intersected by dispersive bands then it can exhibit the discontinuity of Hilbert–Schmidt distance between eigenmodes corresponding to the wave vectors just before and just after the crossing. Such an effect is called singular band touching \cite{Rhim_2021}. This limiting value of Hilbert–Schmidt distance is bulk invariant, different from the Chern number. 

One of the motivations for the photonic implementation of systems with flat, or actually nearly flat bands \cite{vicencio_poblete_photonic_2021}, was the desire to reduce the group velocity of light in order to compress light in space, which leads to the concentration of the optical signal and an increase in the light-matter interaction, or the enhancement of non-linear effects. Another, more obvious application is the possibility of realizing delay lines that can buffer the signal to adjust the timing of optical signals \cite{Baba_2008}.
A promising alternative to photonic circuits are magnonic systems, which allow signals of much shorter wavelengths to be processed in devices several orders of magnitude smaller \cite{Chumak_2022, Mahmoud_2020}.  For this reason, it seems natural to seek a magnonic realization of Lieb lattices. 

In this paper, we propose the realization of such lattices based on a magnonic structure in the form of a perpendicularly magnetized magnetic layer with spatially modulated material parameters or spatially varying static internal field.
Lieb lattices have been studied also in the context of magnetic properties, mainly due to the possibility of enhancing ferromagnetism in systems of correlated electrons \cite{Tasaki_1998}, where the occurrence of flat bands with zero kinetic energy was used to expose the interactions. There are also known single works where the spin waves have been studied in the Heisenberg model in an atomic Lieb lattice, such as the work on the magnon Hall effect \cite{Cao_2015}. But the comprehensive studies of spin waves in nanostructures that realize magnonic Lieb lattices and focus on wave effects in a continuous model have not been carried out so far. In this work, we demonstrate the possibility of realization of magnonic lattices in planar structure based on low spin wave damping material: yttrium iron garnet (YIG) where the iron is partially substituted by gallium~(Ga). We present the dispersion relation with a weakly depressive (flat) band exhibiting the compact localized spin waves. The flat is almost intersected at the 
$M$ point of the 1$^{\rm st}$ Brillouin zone by highly dispersive bands, similar to Dirac cones. We discuss the spin wave spectra and compact localized modes both for simple and extended Lieb lattices.

The introduction is followed by the section describing the model and numerical method we used, which precedes the main section where the results are presented and discussed. The paper is summarized by conclusion and supplemented with additional materials where we showed: (A) the results for extended Lieb-7 lattice, (B) an alternative magnonic Lieb lattice design via shaping the demagnetizing field, and (C) an attempt of formation magnonic Lieb lattice by dipolarly coupled magnetic nanoelements, (D) discussion of small differences in the demagnetizing field of majority and minority nodes responsible for opening a small gap in the Lieb lattice spectrum.


\section{\label{sec:level1} Structure }

Magnonic crystals (MCs) are regarded as promising structures for magnonic-based device applications \cite{Krawczyk_2014, Mahmoud_2020}. In our studies, we consider planar MCs to design the magnonic Lieb lattice, owing to the relative ease of fabrication of such structures and their experimental characterization \cite{Choudhury_2016,Tacci_2012,Gubbiotti_2012}. We proposed realistic systems that mimic the main features of the tight-binding model of Lieb lattice \cite{Beugeling_2012,Rhim_2021}.

\begin{figure}[!ht]
\vspace{0.5cm}
\includegraphics[width=8cm]{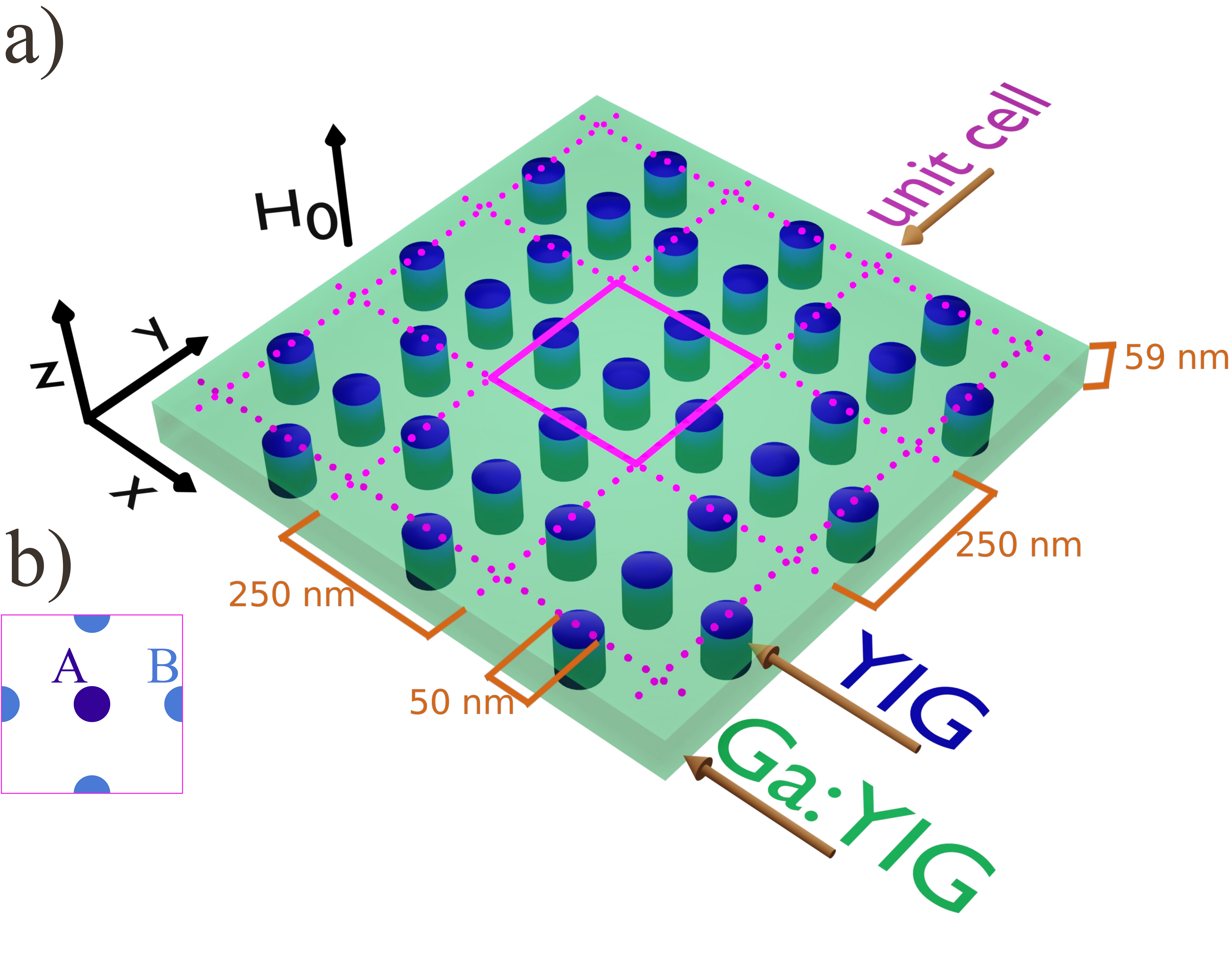}
   \caption{\small{Basic magnonic Lieb lattice. The planar magnonic structure consists of YIG cylindrical nanoelements embedded within Ga:YIG. Dimensions of the ferromagnetic unit cell are equal to 250x250x59 nm and the unit cell contains three inclusions of 50 nm diameter. (a) The structure of basic Lieb lattice, and (b) top view of the Lieb lattice unit cell where the node (inclusion) from minority sublattice $A$ and two nodes (inclusions) from two majority sublattices $B$ are marked.} }
\label{Fig:structure1}
\end{figure}

Investigated MCs consist of yttrium iron garnet doped with gallium (Ga:YIG) matrix and yttrium iron garnet (YIG) cylindrical inclusions arranged in Lieb lattice Fig.~\ref{Fig:structure1}. Doping YIG with Gallium is a procedure where magnetic ${\rm Fe}^{3+}$ ions are replaced by non-magnetic ${\rm Ga}^{3+}$ ions. This method not only decreases saturation magnetization $M_{S}$ but, simultaneously, arises uni-axial out-of-plane anisotropy, that ensures the out-of-plane orientation of static magnetization in Ga:YIG layer at a relatively low external field applied perpendicularly to the layer. Discussed geometry, i.e. forward volume magnetostatic spin wave configuration, does not introduce an additional anisotropy in the propagation of spin waves, related to the orientation of static magnetization. 

The design of the Lieb lattice requires the partial localization of spin wave in inclusions, which can be treated as an approximation of the nodes from the tight-binding model. Furthermore, the neighbouring inclusions in the lattice have to be coupled strongly enough to sustain the collective spin wave dynamics, and weakly enough to minimize the coupling between further neighbours. Therefore, the geometrical and material parameters were selected to ensure the occurrence of oscillatory excitations in the (YIG) inclusions and exponentially evanescent spin waves in the (Ga:YIG) matrix. The size of inclusions was chosen small enough to separate  three lowest magnonic bands with almost uniform magnetization precession inside the inclusion from the bands of higher frequency, where the spin waves are quantised inside the inclusions. Also, the thickness of the matrix and inclusion was chosen in a way that there are no nodal lines inside the inclusion. The condition which guarantee the focussing magnetization dynamics inside the inclusions is fulfilled in the frequency range below the ferromagnetic resonance (FMR) frequency of the out-of-plane magnetized layer made of Ga:YIG (matrix material): $f_{\rm FMR,Ga:YIG}=$4.95~GHz and above the FMR frequency of out-of-plane magnetized layer made of YIG (inclusions material): $f_{\rm FMR,YIG}=2.42$~GHz. These limiting values were obtained using the Kittel formula for out-of-plane magnetised film: $f_{\rm FMR}=\frac{\gamma}{2\pi}\text{\textbar}{\mu_0H_{\rm 0}+\mu_0H_{\rm ani}-\mu_0M_{S}\text{\textbar}}$, where we used the following values of material parameters \cite{Bottcher_2022} for YIG: gyromagnetic ratio $\gamma = 177~\frac{\rm GHz}{\rm T}$, magnetization saturation $\mu_{0}M_{ S} = 182.4$~mT, exchange stiffness constant $A = 3.68~\frac{\rm pJ}{\rm m}$, (first order) uni-axial anisotropy field $\mu_{0}H_{\rm ani} = -3.5~$mT, and for Ga:YiG: $\gamma = 179~\frac{\rm GHz}{\rm T}$, $\mu_{0}M_{ S} = 20.2$~mT, $A = 1.37~\frac{\rm pJ}{\rm m}$, $\mu_{0}H_{\rm ani} = 94.1$~mT.
Since the greatest impact of the first order uniaxial anisotropy field ($\mu_{0}H_{\rm ani}$), we decided to neglect higher order terms of uni-axial anisotropy and cubic anisotropy of (Ga:)YIG. Due to the presence of out-of-plane anisotropy and relatively low saturation magnetization, we could consider a small external magnetic field $\mu_{0}H_{0} = 100$~mT to reach saturation state. 

\begin{figure}[!ht]
\vspace{0.5 cm}
\includegraphics[width=8cm]{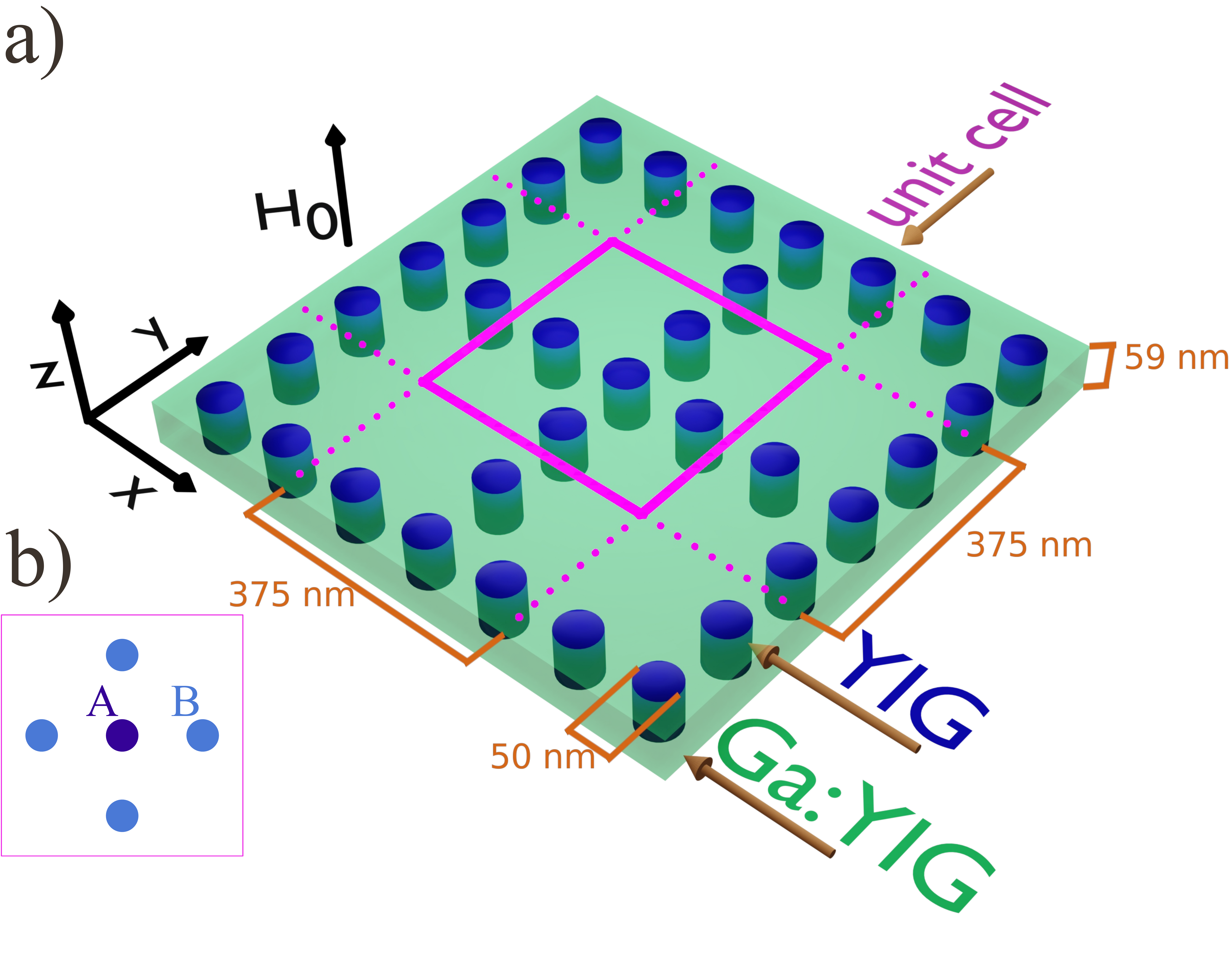}
   \caption{\small{Extended magnonic Lieb lattice -- Lieb~-~5. Dimensions of the unit cell are 375x375x59 nm and contain 5 inclusions of size 50 nm in diameter. Also, we maintain the same separation (distance between centres of neighbouring sites is 125 nm) as for considered basic Lieb lattice -- Fig.~\ref{Fig:structure1}. (a) The structure of Lieb-5 lattice, and (b) top view on Lieb-5 lattice unit cell where the node (inclusion) from minority sublattice $A$ and four nodes (inclusions) from two majority sublattices $B$ are marked.} }
\label{Fig:structure2}
\end{figure}

It is worth noticing that without the evanescent spin waves in the ferromagnetic matrix, the appropriate coupling between inclusions would not be possible. Therefore the realization of the Lieb lattice in form of the array of ferromagnetic nanoelemets embedded in air/vacuum seems to be very challenging --  see the exemplary results in Supplementary Information C. 

We also tested the possibility of other realizations of magnonic Lieb lattices. One solution seemed to be the design of a structure in which the concentration of the spin wave amplitude in the Lieb lattice nodes would be achieved through an appropriately shaped profile of the static demagnetizing field -- Supplementary Information B. However, the obtained results were not as promising as for YIG/Ga:YIG system. 

In the main part of the manuscript, we present the results for the basic Lieb lattice (showed in Fig.~\ref{Fig:structure1}) and extended Lieb-5 lattice (showed in Fig.~\ref{Fig:structure2}), based on  YIG/Ga:YIG structures. 
The further extension of the Lieb lattice may be realized by increasing the number of $B$ nodes between neighbouring $A$ nodes. Supplementary Information A presents the results for Lieb-7, where for each site (inclusion) from minority sublattice $A$, we have six nodes (inclusions), grouped in three-element chains, from majority sublattices $B$. 


\section{\label{sec:level1} Methods }

The spin waves spectra and the spatial profiles of their eigenmodes were obtained numerically in a semi-classical  model, where the dynamics of magnetization vector $\textbf{M}(\textbf{r},t)$ is described by the Landau-Lifshitz equation \cite{Gurevich1996}:
\begin{equation}\label{eq:LLE}
\frac{d\textbf{M}}{dt}=-\gamma\mu_{0}[ \textbf{M} \times \textbf{H}_{\rm eff} + \frac{\alpha}{M_{S}} \textbf{M} \times (\textbf{M} \times \textbf{H}_{\rm eff})].
\end{equation}
The symbol $\textbf{H}_{\rm eff}(\textbf{r},t)$ denotes effective magnetic field.

 In numerical calculations, we neglected the damping term since $\alpha$ is small both for YIG and for YIG with Fe substituted partially by Ga  (for $\alpha_{\rm Ga:YIG} = \num{6.1e-4}$ and $\alpha_{\rm YIG} = \num{1.3e-4}$ \cite{Bottcher_2022}). The effective magnetic field $H_{\rm eff}$ contains the following components: the external field  $H_{0}$, exchange field $H_{\rm ex}$, bulk uniaxial anisotropy field $H_{\rm ani}$ and dipolar field $H_{\rm d}$:
\begin{equation}\label{eq:Heff}
\textbf{H}_{\rm eff}(\textbf{r},t)=H_{0}\vu{z}+\frac{2A}{\mu_{0}M^{2}_{S}}\laplacian\textbf{M}(\textbf{r},t)+H_{\rm ani}(\textbf{r})\vu{z}-\grad\varphi(\textbf{r},t),
\end{equation}
where the $z-$direction is normal to the plane of the magnonic crystal. We assume that the sample is saturated in $z-$direction and magnetization vector precesses around this direction. The material parameters ($M_{S}$, $A$, $\alpha$ and $\gamma$) are constant within matrix and inclusions. 
 
 Using the magnetostatic approximation the dipolar term of the effective magnetic field may be expressed as a gradient of magnetic scalar potential:
\begin{equation}\label{eq:Hd}
\textbf{H}_{\rm d}(\textbf{r},t)=-\grad\varphi(\textbf{r},t)
\end{equation}

By using the Gauss equation magnetic scalar potential may be associated with magnetisation as follows:
\begin{equation}\label{eq:GaussMagnetic}
\laplacian\varphi(\textbf{r},t) = \div\textbf{M}(\textbf{r},t)
\end{equation}

Spin-wave dynamics is calculated numerically using the finite-element method (FEM). We used the COMSOL Multiphysics \cite{Dechaumphai_2019} to implement the Landau-Lifshitz equation (Eq.~\ref{eq:LLE}) and performed FEM computation for the defined geometry of magnonic Lieb lattices.
The COMSOL Multiphysics is the software used for solving a number of physical problems, since many implemented modules it becomes more and more convenient. Nevertheless, all the equations were implemented in the Mathematics module which contains different forms of partial differential equations. Eq.~\ref{eq:LLE} was solved by using eigenfrequency study, on the other hand, to solve Eq.~\ref{eq:GaussMagnetic} we used stationary study. To obtain free decay of scalar magnetic potential in the model we applied $5~\mu$m of a vacuum above and underneath the structure. At the bottom and top surface of the model with vacuum, we applied the Dirichlet boundary condition. We use the Bloch theorem for each variable (magnetostatic potential and components of magnetization vector) at the lateral surfaces of a unit cell. We selected the symmetric unit cell with minority node $A$ in the centre to generate a symmetric mesh which does not perturb the four-fold symmetry of the system -- this approach is of particular importance for the reproduction of the eigenmodes profiles in high-symmetry points. In our numerical studies, we used 2D wave vector $\textbf{k}=k_{x}\hat{\textbf{x}}+k_{y}\hat{\textbf{y}}$ as a parameter for eigenvalue problem which was selected along the high symmetry path $\Gamma-X-M-\Gamma$ to plot the dispersion relation. We considered the lowest 3, 5 and 7 bands for basic Lieb lattice, Lieb-5 lattice and Lieb-7 lattice, respectively.


\section{\label{sec:level1} Results }

The tight-biding model of the basic Lieb lattice, with hopping restricted to next-neighbours gives three bands in the dispersion relation. The top and bottom bands are symmetric with respect to the second, perfectly flat band, and intersect with this dispersionless band at $M$ point of 1$^{\rm st}$ Brillouin zone, with constant slope forming two Dirac cones\cite{Taie_2015,Leykam_2018}. In a realistic magnonic system, the spin wave spectrum showing the particle-hole symmetry with a zero energy flat band is difficult to reproduce because (i) the dipolarly dominated spin waves, propagating in magnetic film, experience a significant reduction of the group velocity with an increase of the wave vector (this tendency is reversed for much larger wave vectors were the exchange interaction starts to dominate) \cite{Gurevich1996}, (ii) the dipolar interaction is long-range. The first effect makes the lowest band wider than the third band, and the latter one -- induces the finite width of the second band \cite{Beugeling_2012}. We are going to show, that this weakly dispersive band supports the existence of CLS. Therefore, we will still refer to it as {\em flat band}, which is a common practice for different kinds of realization of Lieb lattices in photonics or optical lattices.

\begin{figure}[!ht]
\includegraphics[width=8cm]{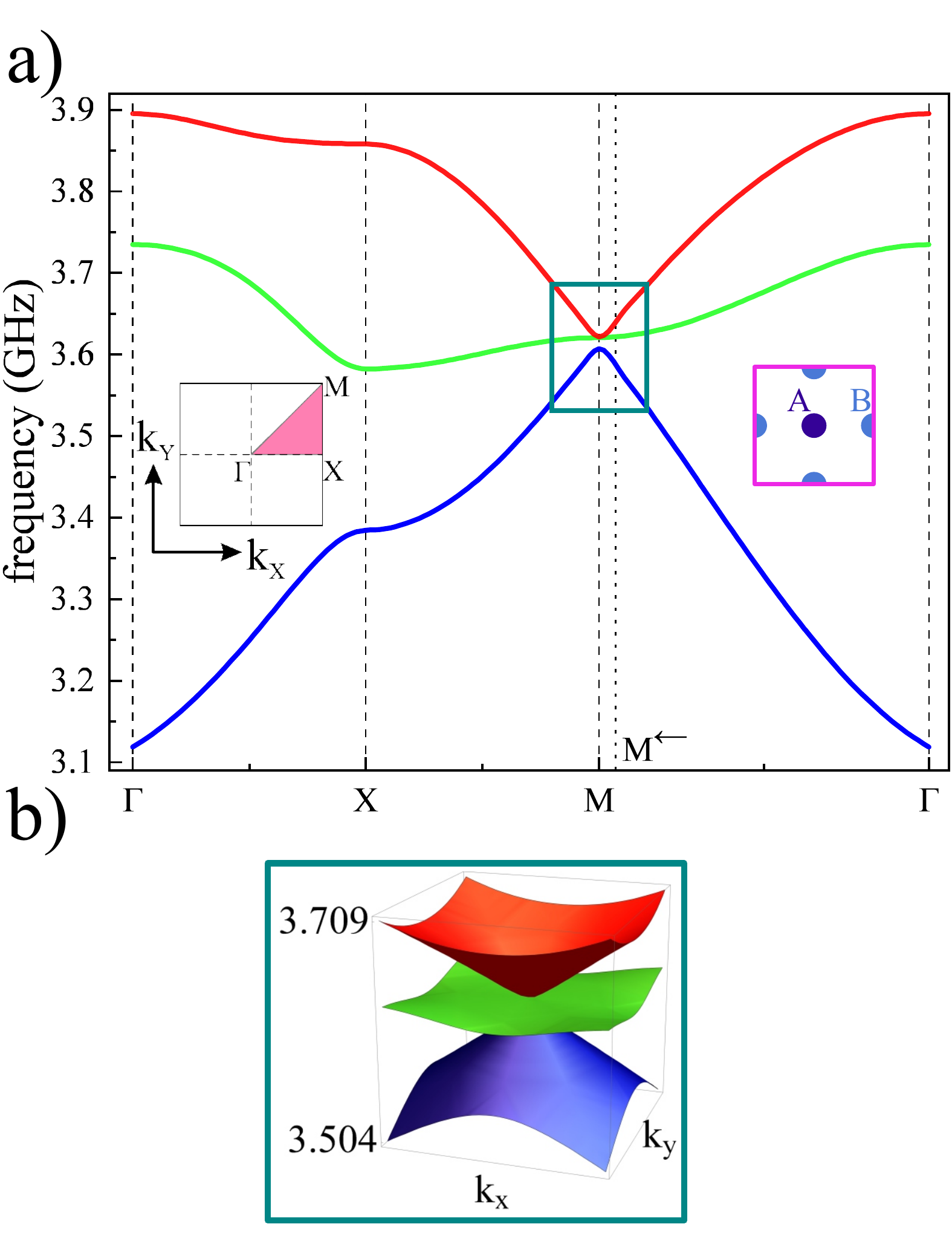}
   \caption{\small{Dispersion relation for the basic magnonic Lieb lattice, containing three inclusions in the unit cell: one inclusion $A$ from minority sublattice and two inclusions $B$ from majority sublattices (see Fig.~\ref{Fig:structure1}). (a) The dispersion relation is plotted along the high symmetry path $\Gamma$-X-M-$\Gamma$ (see the inset). The lowest band (blue) and the highest band (red) create Dirac cones almost touching (b) in the M point. The middle band (green) is relatively flat in the vicinity of the M point. 
   } }
\label{Fig:results_1a}
\end{figure}

The results obtained for the basic magnonic Lieb lattice, (Fig.~\ref{Fig:structure1}), are shown in Fig.~\ref{Fig:results_1a}. As we predicted, three lowest bands form a band structure which is similar to the dispersion relation known from the tight-binding model \cite{Zhang_2017}. However, in a considered realistic system there is an infinite number of higher bands, not shown in  Fig.~\ref{Fig:results_1a}(a). For higher bands, spin waves can propagate in an oscillatory manner in the matrix hence the system does not mimic the Lieb lattice where the excitations should be associated with the nodes (inclusions) of the lattice.

Due to the fourfold symmetry of the system, the dispersion relation could be inspected along the high symmetry path $\Gamma-X-M-\Gamma$. Frequencies of the first three bands are in the range $f_{\rm FMR,YIG}~-~f_{\rm FMR,Ga:YIG}$. Their total width is about $\approx0.78$~GHz. The first and third band form Dirac cones at $M$ point, separated by a tiny gap $\approx15$~MHz. The possible mechanism responsible for opening the gap is a small difference in the demagnetizing field in the areas of inclusions $A$ (from the minority lattice) and inclusions $B$ (from two majority sublattices) -- see Supplementary Information D. Inclusions $A$ ($B$) have four (two) neighbours of type $B$ ($A$). Although inclusions $A$ and $B$ have the same size and are made of the same material, the static field of demagnetization inside them differs slightly due to the different neighbourhoods. This effect is equivalent to the dimerization of the Lieb lattice by varying the energy of the nodes in the tight-binding model, which leads to the opening of a gap between Dirac cones and parabolic flattening of them in very close proximity to the $M$-point. It is worth noting that in the investigated system, the gap opens between the first and second bands, while the second and third bands remain degenerated at point $M$, with numerical accuracy.

The middle band can be described as weakly dispersive. The band is more flat on the $X-M$ path and, in particular, in the vicinity of $M$ point -- see Fig.~\ref{Fig:results_1a}(b). The small width of the second band can be attributed to long-range dipolar interactions which govern the magnetization dynamics in a considered range of sizes and wave vectors. It is known that even the extension of the range of interactions to next-nearest-neighbours in the tight-binding model induces the finite width of the flat band for the Lieb lattice.

\begin{figure}[!ht]
\includegraphics[width=8cm]{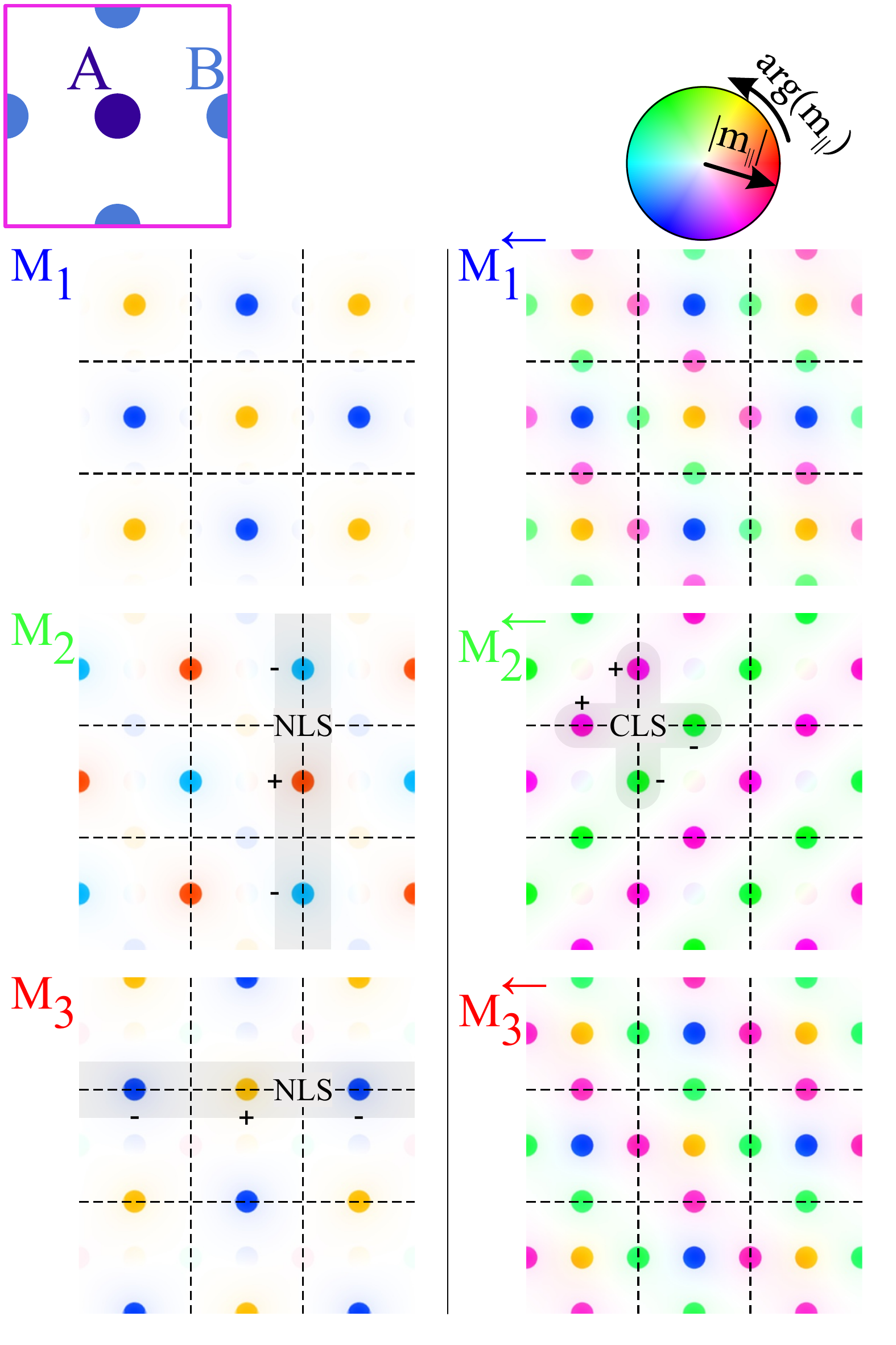}
   \caption{\small{The spin wave profiles obtained for the basic magnonic Lieb lattice, composed of three inclusions in the unit cell (see Fig.~\ref{Fig:structure1}). The modes are presented for each band exactly at $M$ (left column) and in its proximity ($M^\leftarrow$) on the path $M-\Gamma$ (right column).
   In the presented profiles, the saturation and the colour denote the amplitude and phase of the dynamic, in-plane component of magnetization.
   The compact localized states (CLS) are presented at the point $M^\leftarrow$ for the second band -- right column. The CLS do not occupy minority sublattice $A$. The inclusions $B$, in which the magnetization dynamics is focused, are quite well isolated from each other. One can easily notice that the lattice is decorated by loops (marked by grey patches) where the phase of the precessing magnetization flips between inclusions ($+$ and $-$ signs). Exactly at point $M$ -- left column, we observe the degeneracy of the second and third bands. The spin waves occupy $B$ inclusions only in one majority sublattice, i.e. along vertical or horizontal lines, filliping the phase from inclusion to inclusion which gives the pattern characteristic to noncontractible loop states (NLS) - marked by grey stripes. 
}}
\label{Fig:results_1b}
\end{figure}

To prove that the second band supports the CLS regardless of its finite width, we plotted the profiles of spin wave eigenmodes at $M$ point and in its close vicinity. The results are presented in Fig.~\ref{Fig:results_1b}. The profiles were shown for infinite lattice and are presented in the form of square arrays containing 3x3 unit cells, where the dashed lines mark their edges. It is visible that the spin waves are concentrated in the cylindrical inclusions, where the amplitude and phase of precession is quite homogeneous. In calculations, we used the Bloch boundary conditions applied for a single unit cell, which means that at $M$ point the Bloch function is flipped after translation by lattice period, in both principal directions of the lattice and we will not see the single closed loops of CLS or lines of NLS. Exactly at $M$ point, all three bands have zero group velocity. Therefore, the corresponding modes (left column) are not propagating. The lowest band ($M_1$) occupy only inclusions $A$ from the minority sublattice where the static demagnetizing field is slightly lower than inside inclusions $B$ (see Supplementary Information D), which justifies its lower frequency and lifting the degeneracy with two higher modes $M_2$ and $M_3$ of the same frequency. Each of the modes $M_2$ and $M_3$ occupy only one of two sublattices $B$, therefore they can be interpreted as NLS. To observe the pattern typical for CLS, we need to move slightly away from $M$ point. The first and third modes have then the linear dispersion with high group velocity and the second band remains flat. We selected the point $M^\leftarrow$ shifted from $M$ point toward $\Gamma$ point by 5\% of $M-\Gamma$ distance (right column). We can see that the first and third modes $M^\leftarrow_1$, $M^\leftarrow_3$ occupy now all inclusions and the mode $M^\leftarrow_2$ from the flat band has a profile typical for CLS, predicted by tight-binding models \cite{Leykam_2012,Leykam_2018,Chen_2016,Bhattacharya_2019,Zhang_2017}:
\begin{equation}
\text{\textbar}m_\textbf{k}>=[\underbrace{-e^{i\frac{k_{y}}{2}}}_B,\underbrace{0}_A,\underbrace{e^{i\frac{k_{x}}{2}}}_B]\label{eq:CLS_profile}
\end{equation}
where $m_{\textbf{k}}$ is the complex amplitude of the Bloch function in the base of unit cell (i.e. on two inclusion $B$ from majority sublattices and one inclusion $A$ from minority sublattice), 
$\textbf{k}=[k_x,k_y]$ is dimensionless wave vector. From (Eq.~\ref{eq:CLS_profile}), we can see that (i) CLS do not occupy the minority nodes $A$ and (ii) close to $M$ point the phases at two nodes $B$, from different majority sublattices, are opposite. These two features are reproduced for $M^\leftarrow_2$ mode in investigated magnonic Lieb lattice. In the profile of this mode, we marked (by a grey patch) the elementary loop of CLS which is easily identified in finite systems. Here, in an infinite lattice with Bloch boundary conditions, the loops are infinitely replicated with $\pi$ phase shift after each translation $x-$ and $y-$direction. The localization at the inclusions $B$ and the absence of the spin wave dynamics in inclusions $A$ is observed regardless of the wave vector. Therefore, the coupling can take place only between the next neighbours (inclusions $B$), i.e. on larger distances and mostly due to dipolar interactions, that makes the second band not perfectly flat.

\begin{figure}[!t]
\includegraphics[width=8cm]{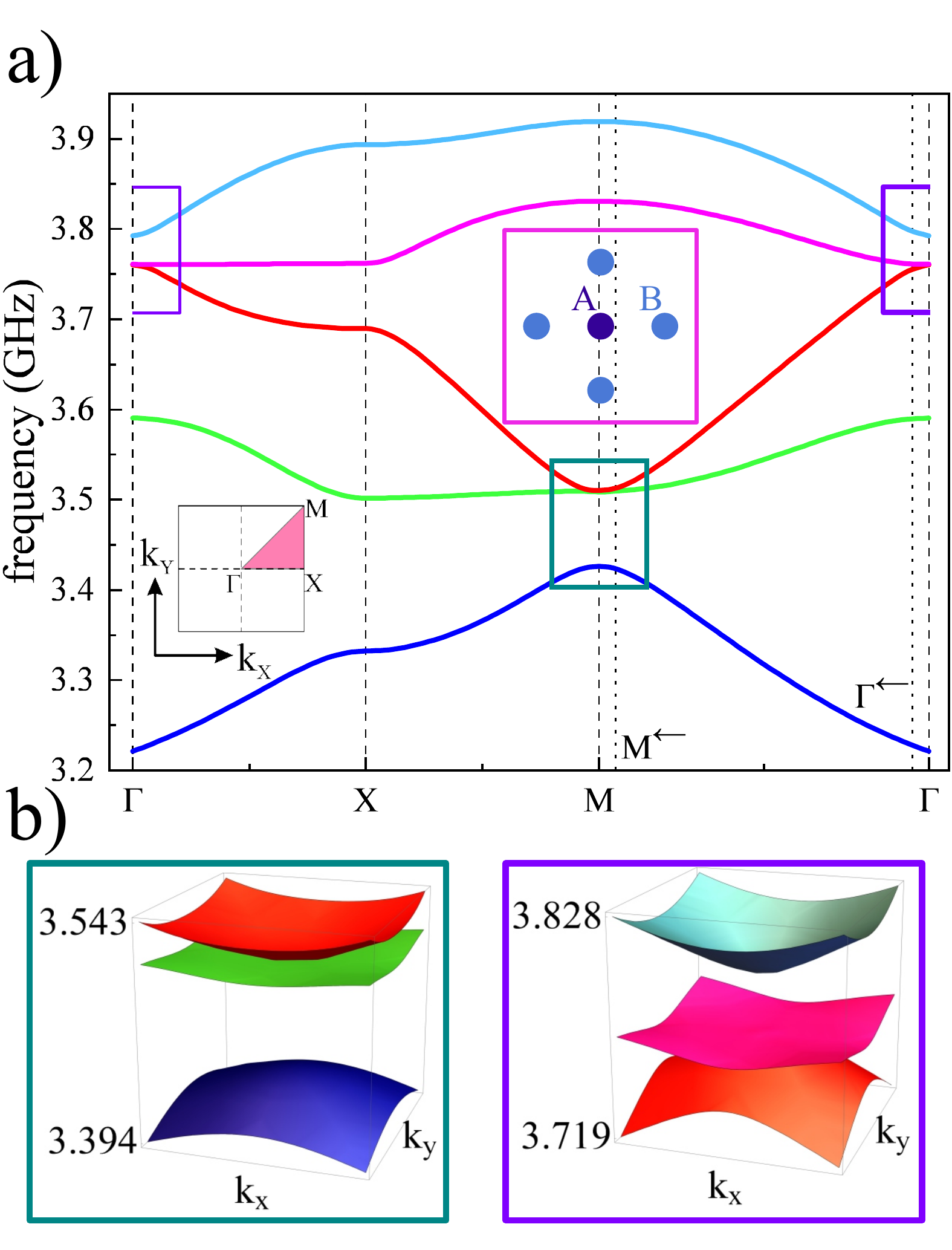}
   \caption{\small{Dispersion relation for the extended magnonic Lieb lattice Lieb-5, containing five inclusions in the unit cell: one inclusion $A$ from minority sublattice and four inclusions $B$ from majority sublattices (see Fig.~\ref{Fig:structure2}). (a) The dispersion relation is plotted along the high symmetry path $\Gamma$-X-M-$\Gamma$ (see the inset).  The first, third and fifth bands  (dark blue, red and cyan) are strongly dispersive bands, while the second and fourth bands (green and magenta)  are less dispersive and related to the presence of CLS. The system does not support the appearance of Dirac cones, even in case when the interaction is fictitiously limited only to inclusions, according to tight-binding model. (b) The zoomed regions in the vicinity of $\Gamma$ (in dark green frame) and $M$ points show the essential gaps with relatively low, parabolic-like curvatures for top and bottom bands.} }
\label{Fig:results_2a}
\end{figure}

Let's discuss now the presence of flat bands and CLS in an extended magnonic Lieb lattice (Lieb-5), containing five inclusions in the unit cell: one
inclusion $A$ form minority sublattice and four inclusions $B$
from majority sublattices, as it is presented in Fig.~\ref{Fig:structure2}. In the considered structure, we add two additional inclusions $B$ into the unit cell in such a way that neighbouring inclusions $A$ are linked by the doublets of inclusions $B$. The sizes of inclusions, distances between them, the thickness of the layer and the material composition of the structure remained the same as for the basic Lieb lattice, discussed earlier (Fig.~\ref{Fig:structure1}).

The dispersion relation obtained for the magnonic Lieb-5 lattice can be found in Fig.~\ref{Fig:results_2a}(a). The properties of the extended Lieb lattices are well described in the literature \cite{Zhang_2017,Dias_2015,Liu_2020,Mao_2020}. The tight-binding model description of Lieb-5 lattices, with information about their dispersion relation and the profiles of the eigenmodes, are presented in numerous papers\cite{Zhang_2017,Leykam_2012,Rhim_2021}. Therefore, it is possible to compare the obtained results with the theoretical predictions of the tight-binding model.

\begin{figure*}[!ht]
\includegraphics[width=\textwidth,height=\textheight,keepaspectratio]{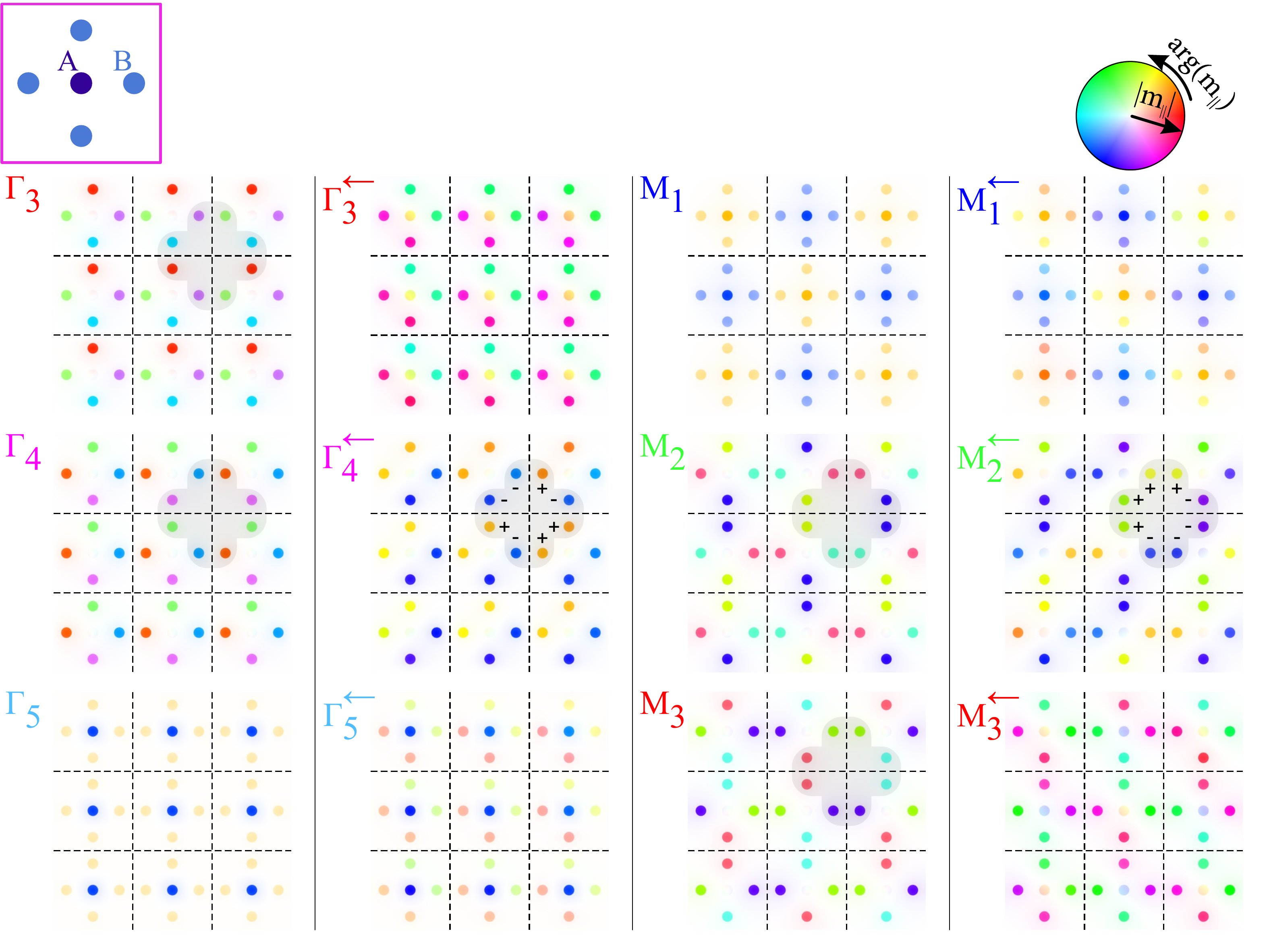}
\caption{\small{The profiles obtained for the extended Lieb lattice consisted of 5 inclusions in the unit cell. The modes are presented for bands No. 3-5 in $\Gamma$ point and its proximity $\Gamma^{\leftarrow}$ (the first and second column). In the third and fourth columns, we presented the profiles for bands No. 1-3 at $M$ point and its vicinity $M^{\leftarrow}$. Each profile of eigenmode is presented on a grid composed of 3x3 unit cells - dashed lines mark the edges of unit cells. The scheme of the unit cell is presented in top-left corner. Exactly at $\Gamma$ (and $M$) point the bands No. 3 and 4 (No. 2 and 3) are degenerated and profiles: $\Gamma_3$ and $\Gamma_4$ ($M_2$ and $M_3$) have non-standard (for CLS) complementary form -- i.e. their combinations $\Gamma_3\pm i\Gamma_4$ ($M_2\pm i M_3$) gives NLS. To obtain proper profiles of CLS, where the phase of procession flips around CLS loop, we need to explore the vicinity of $\Gamma$ ($M$) point -- see the grey patches for the mode $\Gamma^{\leftarrow}_4$ ($M^{\leftarrow}_2$) with $+$ and $-$ signs.} }
\label{Fig:results_2b} 
\end{figure*}

The tight-binding model of Lieb-5 lattice predicts two flat bands with CLS: the second (green) and fourth (magenta) band in the spectrum. The flat bands in the tight-binding model are not intersected by Dirac cones but they are degenerated at $\Gamma$ and $M$ point with the third band (red). These features are reproduced in investigated magnonic Lieb-5 lattice (Fig.~\ref{Fig:structure2}). The dispersion relation for this system is presented in the Fig.~\ref{Fig:results_2a}(a). Also, we have marked, with two rectangles (dark green and violet), the vicinities of $\Gamma$ and $M$ points, where the flat bands (the fourth and second bands) become degenerated with the third, dispersive band -- Fig.~\ref{Fig:results_2a}(b). It is easy to notice the essential frequency gaps ($\approx33$~MHz and $\approx84~$MHz at $\Gamma$ and $M$ points, respectively), which qualitatively corresponds to the prediction of the tight-binding model. It is worth noting that although the low dispersion bands (the second and fourth band) are in general not perfectly flat. Nevertheless, around the point $\Gamma$ and $M$ points the bands are flattened and the $\Gamma-X$ and $X-M$ sections are very flat for the fourth and second band, respectively. 

The spin wave profiles of CLS at the high symmetry points: $\Gamma$ and $M$ are presented in Fig.~\ref{Fig:results_2b}. Exactly at $\Gamma$ and $M$ (the first and third column), we can see the pairs of degenerated mods $\Gamma_3$, $\Gamma_4$ and $M_2$, $M_3$ which exhibit features of CLS predicated by the tight-binding model (see the loops of sites on grey patches): (i) modes occupy only the inclusions $B$ from majority sublattices, (ii) doublets of inclusions $B$ in the loops of CLS have opposite (the same) phases at $\Gamma$ ($M$) point. The significant difference is that; once we switch one to another $B$-$B$ doublet, circulating the CLS loop the phase of precession charges by $\pm \pi/2$ not by 0 or $\pi$. However, when we make combinations of degenerated modes: $\Gamma_3\pm i\Gamma_4$ or $M_2\pm i M_3$, then we obtain the NLS occupying the horizontal or vertical lines, where the precession at exited $B$ inclusion will be in- or out-of-phase. The CLS modes are clearly visible when we move slightly away from the high symmetry point where the degeneracy occurs. In the proximity of $\Gamma$ and $M$ point, one can see the CLS modes $\Gamma^{\leftarrow}_4$ and $M^{\leftarrow}_2$ for which the phase of precession takes the relative values close to 0 or $\pi$. The small discrepancies, are visible as a slight change in the colours representing the phase, resulting from the fact that we are not exactly in high symmetry points but shifted by 5\% on the path $\Gamma-M$. 

The extension of the presented analysis to magnonic Lieb~-~7 lattice, where the inclusions $A$ are liked by the chains composed of three inclusions $B$, is presented in Supplementary Information A.


\section{\label{sec:level1} Conclusions }

We proposed a possible realisation of the magnonic Lieb lattices where the compact localized spin wave modes can be observed in flat bands. The presented system qualitatively reproduces the spectral properties and the localization features of the modes, predicted by the tight-binding model and observed for photonic and electronic counterparts. The magnonic platform for the experimental studies of Lieb lattices seems to be attractive due to the larger flexibility in designing magnonic systems and the steering of its magnetic configuration by external biases. The idea of the magnonic Lieb latices allows considering many problems related to dynamics, localization and interactions in flat band systems taking the advantage of the magnonic systems: presence and possibility of tailoring of long-range interactions, intrinsic non-linearity, etc.


\section{\label{sec:level1} Acknowledgements }
This work has received funding from National Science Centre Poland grants UMO-2020/39/O/ST5/02110, UMO-2021/43/I/ST3/00550 and support from the Polish National Agency for Academic Exchange grant BPN/PRE/2022/1/00014/U/00001.


\bibliography{APS_biblography}


\clearpage

\section*{\label{sec:level1} Supplementary Information }

\subsection{\label{sec:level2}Double extended Lieb lattice}

\begin{figure}[!h]
\includegraphics[width=8cm]{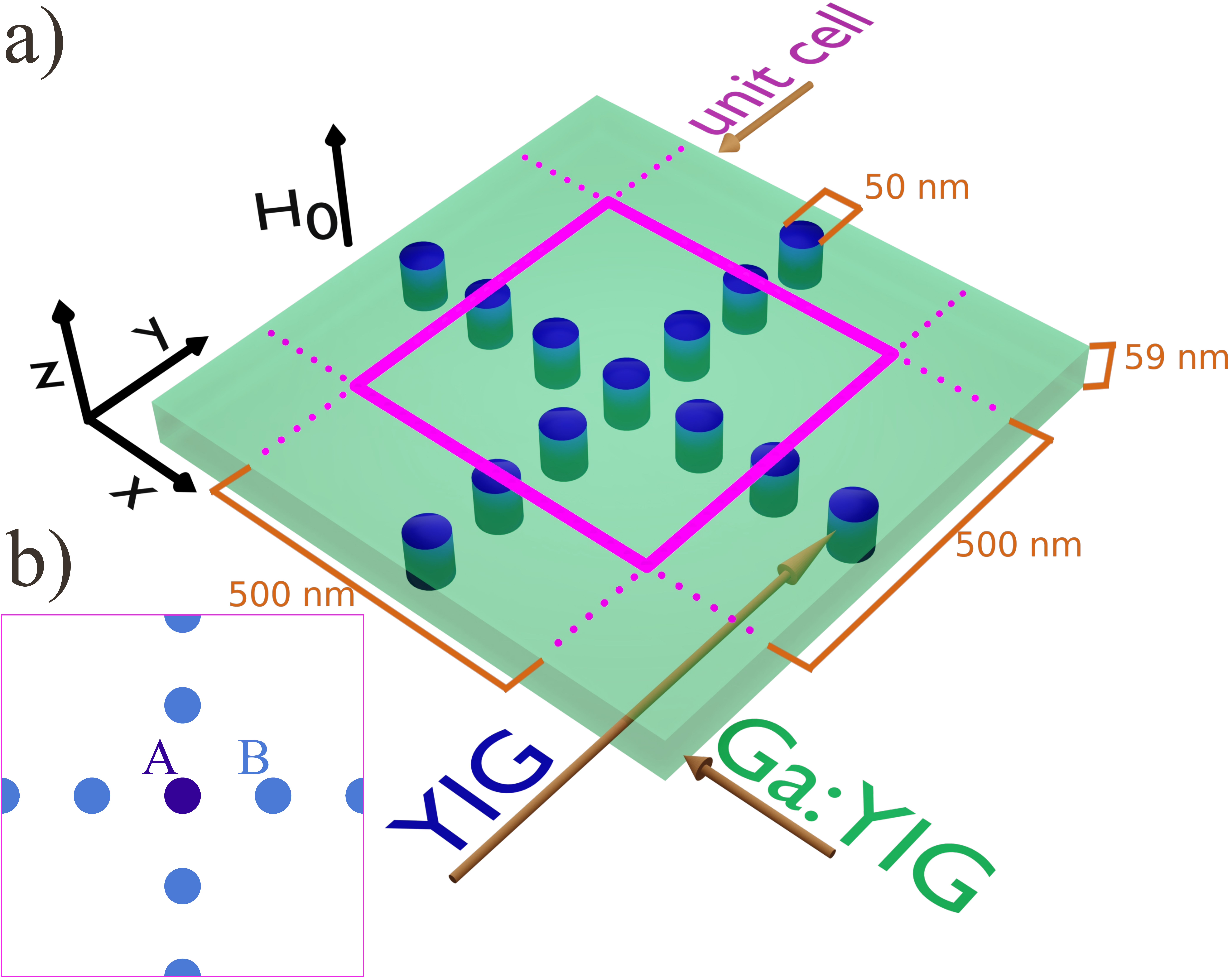}
   \caption{\small{Doubly extended magnonic Lieb lattice: Lieb-7. Dimensions of the ferromagnetic unit cell are equal to 500x500x59 nm. The unit cell contains seven inclusions of 50 nm diameter. (a) The structure of extended Lieb lattice, and  (b) top view on Lieb-7 lattice unit cell where the node (inclusion) from minority sublattice $A$ and two nodes (inclusions) from two majority sublattices $B$ are marked.} }
\label{Fig:structure3}
\end{figure}

We can generate further extensions of the magonic Lieb lattice by adding more inclusions $B$, i.e. by introducing additional majority sublattices.  We considered here a doubly extended Lieb lattice (Lieb-7) to check to what extent the magnonic system corresponds to the tight-binding model. The mentioned lattice consists of seven nodes; six belong to majority sublattices $B$ and one belongs to minority sublattice $A$ (Fig.~\ref{Fig:structure3}). The magnetic parameters were kept as for basic and Lieb-5 lattices, considered in the manuscript. The geometrical parameters have changed only as a result of the introduction of additional inclusions $B$. Therefore, the unit cell has increased to the size 500x500 nm.

In the case of a doubly extended Lieb lattice (Lieb-7), we expect (according to the works \cite{Zhang_2017, Mao_2020}) to obtain seven bands in the dispersion relation. The tight-binding model predicts that the bands will be symmetric with respect to the fourth band, exhibiting particle-hole symmetry. However, due to the dipolar interaction, we did not expect such symmetry. Another feature which one may deduce from the tight-binding model is that bands No. 2, 4 and 6 should be flat while bands No. 1, 3, 5 and 7 are considered dispersive. Moreover, bands No. 3 and 5 suppose to form a Dirac cone intersecting flat band No. 4 at the $\Gamma$ point.

\begin{figure}[!b]
\includegraphics[width=8cm]{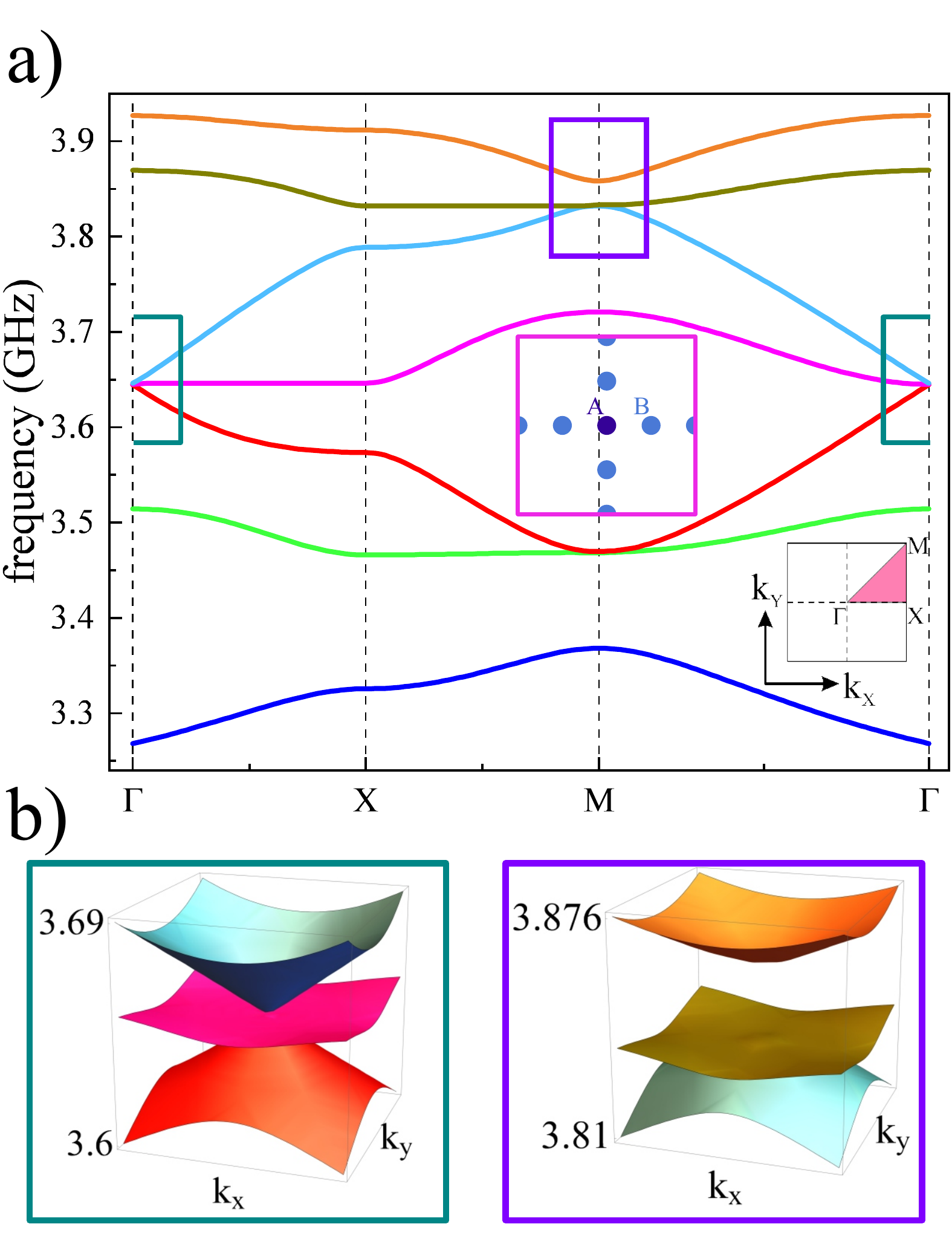}
   \caption{\small{Dispersion relation for the double extended magnonic Lieb lattice (Lieb-7) containing seven inclusions in the unit cell: one inclusion $A$ from minority sublattice and six inclusions $B$ from majority sublattices (see Fig.~\ref{Fig:structure2}). (a) The dispersion relation is plotted along the high symmetry path $\Gamma$-X-M-$\Gamma$ (see the inset). The first, third, fifth and seven bands (dark blue, red, cyan and orange) are dispersive, while the second, fourth and sixth bands (green, magenta and dark green bands) are the flatter bands, supporting the magnonic CLS. Dirac cones occur at the $\Gamma$ point and almost interact with the flatter fourth band, while at $M$ point, we observe the degeneracy of the dispersive parabolic third (fifth) band with a flatter second (six) band. (b) The zoomed vicinity of $\Gamma$ point (dark green frame) and $M$ point (violet frame) regions are presented in 3D. } }
\label{Fig:results_3a}
\end{figure}

\begin{figure}[!ht]
\includegraphics[width=8cm]{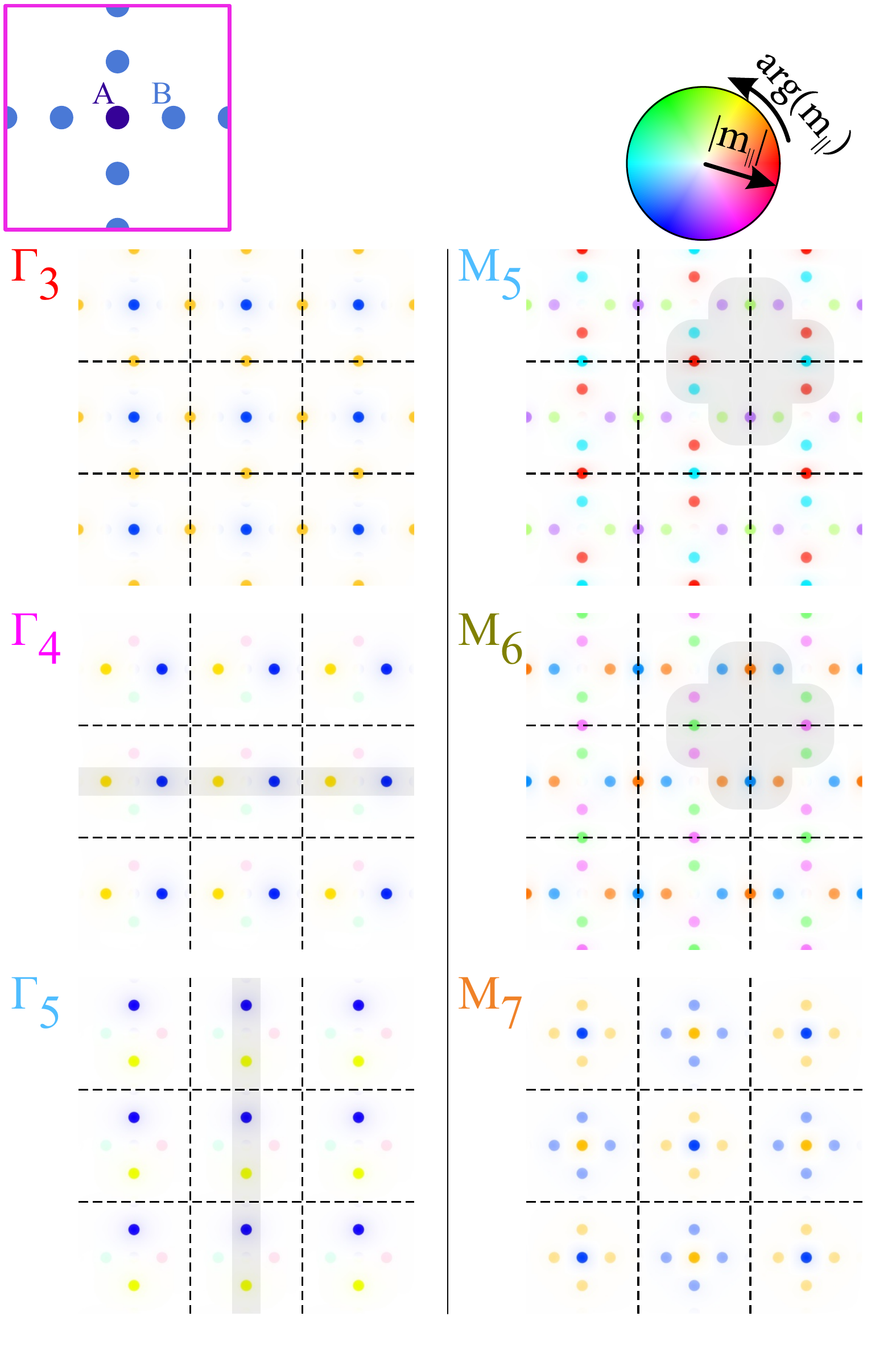}
   \caption{\small{The profiles of eigenmodes were obtained for magnonic Lieb-7. The modes are presented for bands No. 3-5 at $\Gamma$ point and 5-7 at $M$ point. The modes denoted as $\Gamma_{3}$ and $\Gamma_{4}$ are degenerated whereas the $\Gamma_{5}$ is separated from them by extremely small gap $\approx~2$~MHz. At $M$ point, we showed the profiles for bands No. 5, 6 and 7. The modes $M_{5}$ and $M_{6}$ are degenerated and separated from $M_{7}$ by essential gap -- predicted by the tight-binding model.} }
\label{Fig:results_3b}
\end{figure}

We calculated the dispersion relation for magnonic Lieb-7 lattice (Fig.~\ref{Fig:results_3a}(a)), which share many properties with those characteristic for the tight-binding model \cite{Mao_2020}: (i) third and fifth bands form the Dirac cones which almost intersect the flatter forth band at $\Gamma$ point; (ii) the third (and fifth) band has a parabolic shape at $M$ point where it is degenerated with the second (and six) band which is weakly dispersive. The mentioned regions of dispersion are presented as 3D plots in Fig.~\ref{Fig:results_3a}(b). Also, we are going to discuss shortly the profiles of spin wave eigenmodes (including CLS) in these two regions of the dispersion relation, which are presented in Fig.~\ref{Fig:results_3b}. 

Dirac cones appear at the $\Gamma$ point for bands No. 3 and 5. At this point, as for the basic magnonic Lieb lattice (Fig.~\ref{Fig:results_1a}), there is a very narrow gap of the width $\approx~2$~MHz. The profiles $\Gamma_4$ and $\Gamma_5$ (left column in Fig.~\ref{Fig:results_3b}) represent the degenerated states originating from flat and dispersive bands. Both of them do not occupy the inclusions $A$ and are more focused on two inclusions $B$ arranged in horizontal ($\Gamma_4$) and vertical lines ($\Gamma_5$) -- see grey stripes. Therefore, their profiles are similar to NLS, where the first and third inclusion $B$ in each three-element chain, linking inclusions $A$, precesses out-of-phase and the second (central) inclusion $B$ remains unoccupied.

At the $M$ point, the $M_{5}$ and $M_{6}$ bands are degenerated. For these bands, the spin waves are localized in all inclusions $B$ and do not occupy inclusions $A$ (see right column of Fig.~\ref{Fig:results_2b}) The first and third inclusion $B$ in each three-element chain, linking inclusions $A$, precess in-phase, whereas the second (central) inclusion $B$ precesses out-of-phase with respect to the first and third one. This pattern of occupation of inclusions and the phase relations between them is similar to one observed for CLS (see grey patches marking the loops of inclusions in the left column of Fig.~\ref{Fig:results_2b}), but has one significant difference. The phase difference between successive three-element chains of inclusions $B$, in the loop, is equal to $\pm \pi/2$. However, the linear combination of the modes $M_5\pm i M_6$ produces, similarly to the case of the Lieb-5 lattice, the NLS. To observe the proper profiles of CLS or NLS, we need to shift slightly from the high symmetry points $\Gamma$ and $M$ to cancel the degeneracy.

\subsection{\label{sec:level2} Realization of Lieb lattice\\  by shaping demagnetizing field}

\begin{figure}[!ht]
\includegraphics[width=8cm]{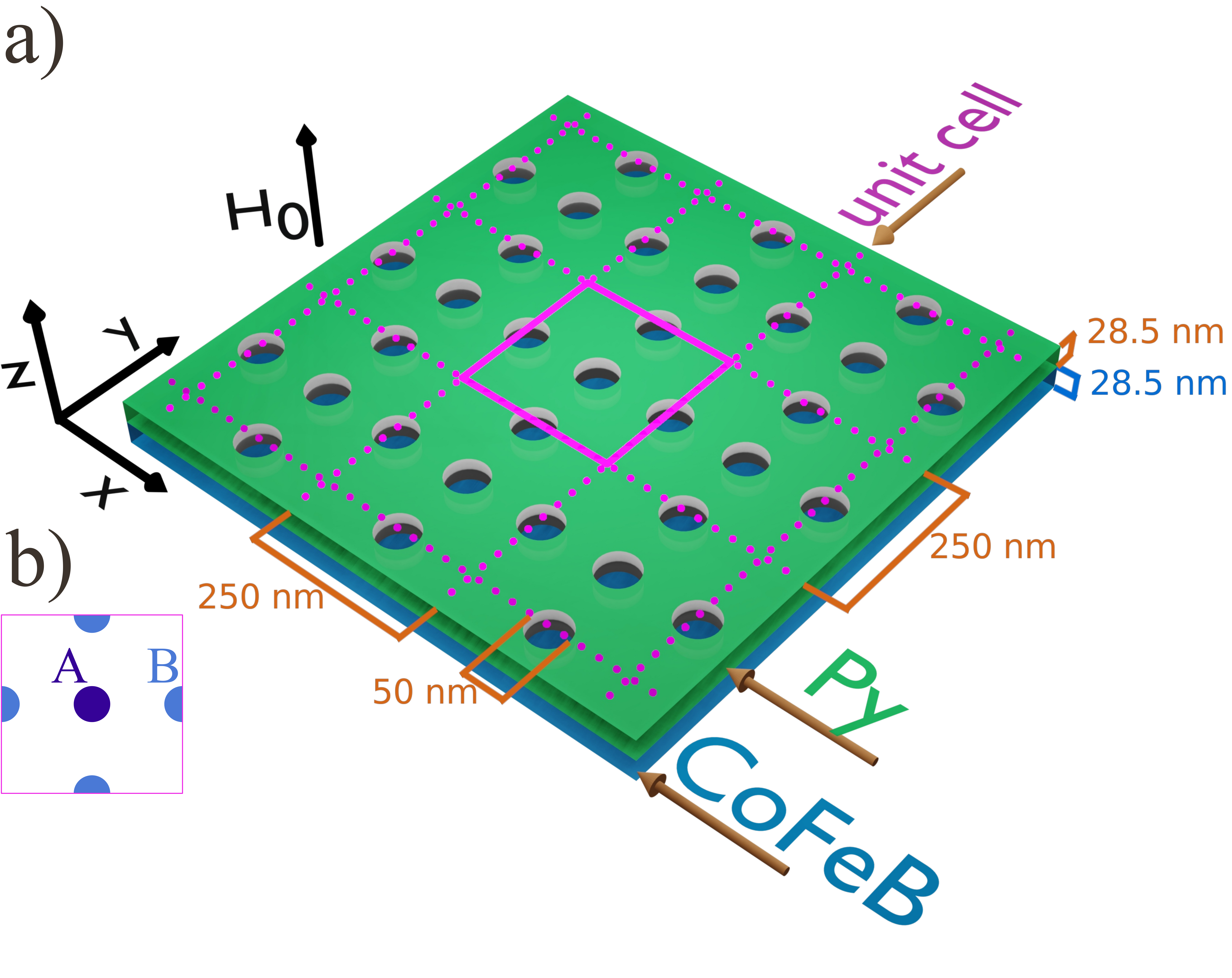}
   \caption{\small{Basic magnonic Lieb lattice where spin wave excitations in the CoFeB layer are shaped by demagnetizing field from Py antidot lattice. Dimensions of the ferromagnetic unit cell are equal to 250x250x59 nm and contain 3 inclusions of 50 nm diameter. (a) structure of basic Lieb lattice, (b) top view on basic Lieb lattice unit cell and differentiation to nodes of sublattice $A$ and $B$.} }
\label{Fig:structure1_demag}
\end{figure}

We have considered also an alternative realisation method for a magnonic Lieb lattice in a ferromagnetic layer. This approach is based on shaping the internal demagnetizing field. The structure under consideration is presented in Fig.~\ref{Fig:structure1_demag}. It consists of a thin (28.5 nm) and infinite CoFeB layer on which a Py antidot lattice (ADL), of 28.5 nm thickness, is deposited. The cylindrical holes in ADL are arranged in shape of the basic Lieb lattice. The size of the unit cell and diameter of holes remains the same as for the basic Lieb lattice proposed in the main part of the manuscript (see Fig.~\ref{Fig:structure1}). Due to the absence of perpendicular magnetic anisotropy (PMA), we decided to apply a much larger external magnetic field ($H_{0}=1500$~mT) to saturate the ferromagnetic material in an out-of-plane direction. 

We assumed the same gyromagentic ratio for both materials $\gamma=187~{\rm \frac{GHz}{T}}$, the following values of material parameters for CoFeB \cite{Graczyk_2017}: saturation magnetization - $M_{S}=1150~{\rm \frac{kA}{m}}$, exchange stiffness constant - $A=15~{\rm \frac{pJ}{m}}$. For Py, we used material parameters  \cite{Gallardo_2014}: saturation magnetization - $M_{S}=796~{\rm\frac{kA}{m}}$, exchange stiffiness constant - $A=13~{\rm \frac{pJ}{m}}$.

The deposition of the ADL made of Py (material of lower $M_{S}$) above the CoFeB layer (material of higher $M_{S}$) is critical for spin wave localization in CoFeB below the exposed parts (holes) of the ADL. The demagnetization field produced on CoFeB/Air interface creates wells partially confining the spin waves. However, this pattern of internal demagnetizing field becomes smoother with increasing distance from the ADL. 

\begin{figure}[!t]
\includegraphics[width=8cm]{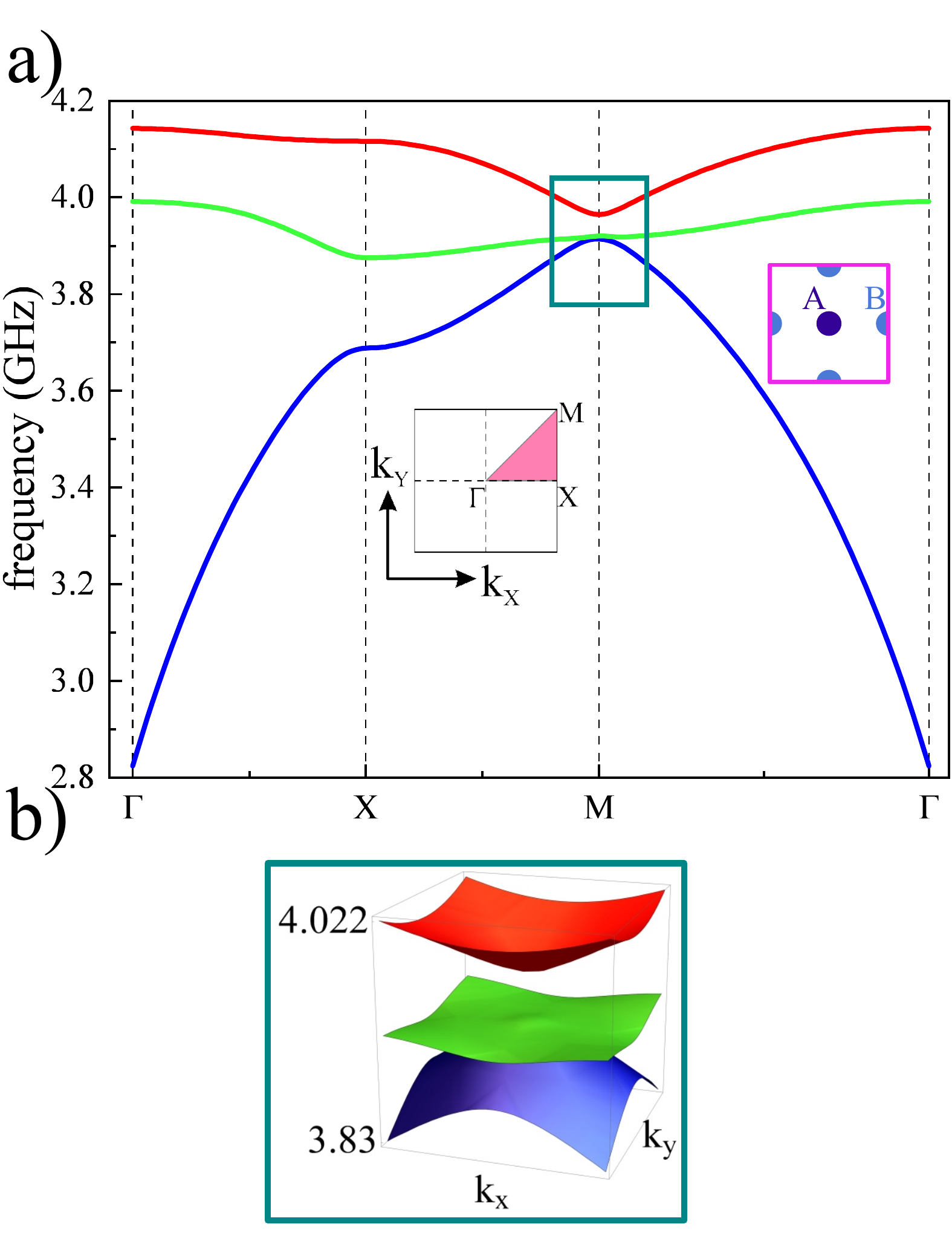}
\caption{\small{The dispersion relation obtained for basic Lieb lattice formed by demagnetizing field of antidot lattice (see Fig.~\ref{Fig:structure1_demag}). (a) The dispersion relation, (b) the 3D plot of dispersion relation in the region marked with the green frame in (a). Results were obtained for $H_{0}=1500$~mT applied out-of-plane. } }
\label{Fig:results_4a}
\end{figure}

The obtained dispersion relation is shown in Fig.~\ref{Fig:results_4a}. It is worth noting that the lowest band is very dispersive, while the highest band is flattened more than in the case of the structure presented in the main part of the manuscript (see Fig.~\ref{Fig:results_1a}). The middle band, which suppose to support CLS, varies in extent similar to the third band. For this structure, Dirac cones in the $M$ point cannot be clearly unidentified.

\subsection{\label{sec:level2}Lieb lattice formed by YIG inclusions \\ in non-magnetic matrix}

The periodic arrangement of ferromagnetic cylinders surrounded by nonmagnetic material (e.g. air) seems to be the simplest realization of the Lieb lattice. To refer this structure to the bi-component system investigated in the main part of the manuscript, we assumed the same material and geometrical parameters for inclusions as for the structure presented in Fig.~\ref{Fig:structure1}.

The advantage of this system is that the confinement of spin waves within the areas of inclusions is ensured for arbitrarily high frequency. We are not limited here by the FMR frequency of the matrix, as it was for bi-component Lieb lattices (Figs.~\ref{Fig:structure1}, \ref{Fig:structure2}). However, the coupling of magnetization dynamics between the inclusions is here provided solely by the dynamical demagnetizing field, i.e. the evanescent spin waves do not participate in the coupling. Therefore, the interaction between inclusions is much smaller in general, which leads to a significant narrowing of all magnonic bands (Fig.~\ref{Fig:results_4a}). The widths of the second and third band can be even smaller than the gap separating from the first bands -- Fig.~\ref{Fig:results_4a}(b). Such strong modification of the spectrum makes the applicability of the considered system for the studies of magnonic CLS questionable.

\begin{figure}[!t]
\includegraphics[width=8cm]{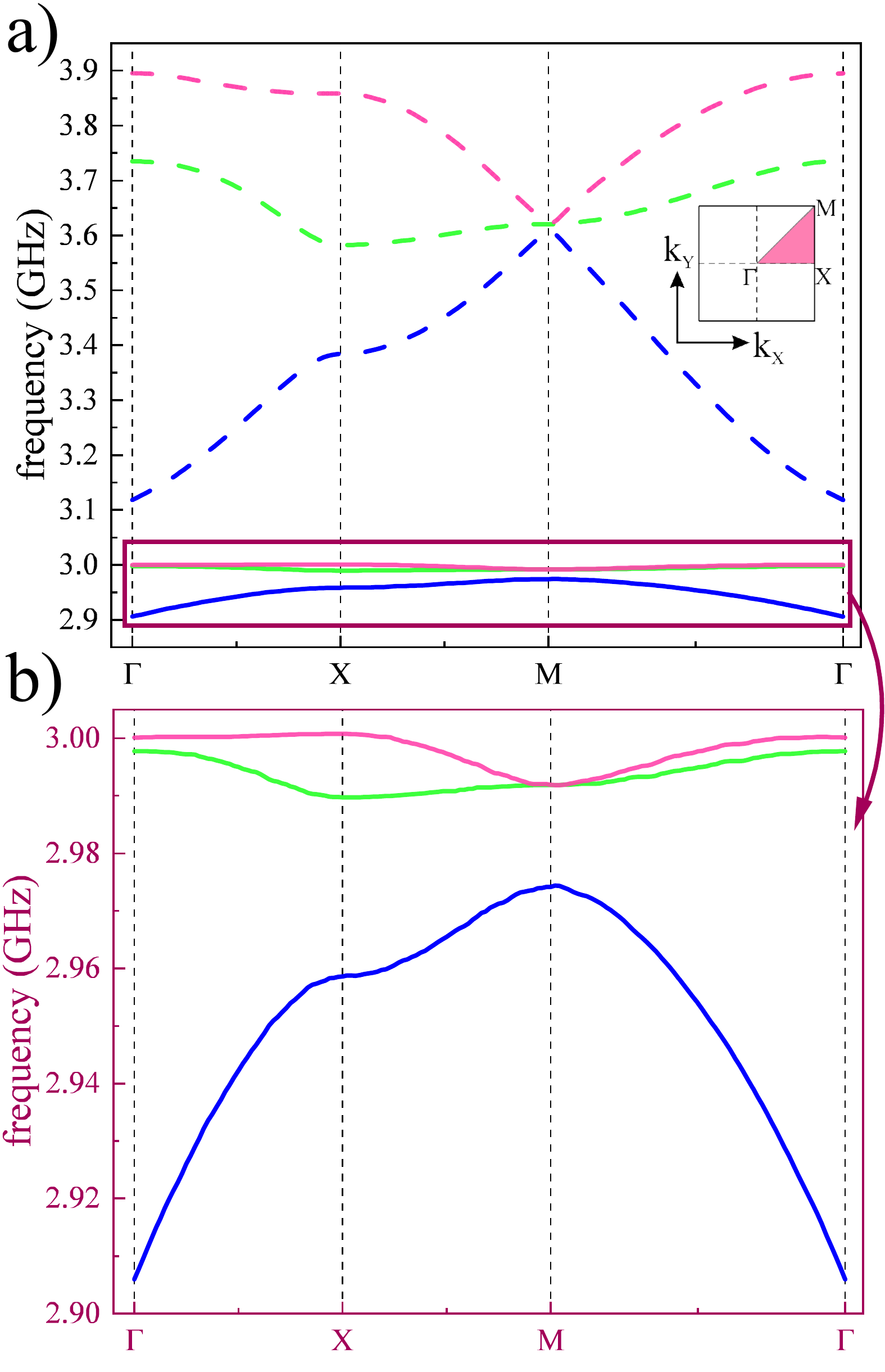}
   \caption{\small{Dispersion relations for basic Lieb lattice. (a) The results obtained for YIG inclusions in Ga:YIG matrix (dashed lines) and YIG inclusions without matrix (solid lines). (b) The zoomed dispersion relation obtained for YIG inclusions without matrix, marked in (a) by the frame. } }
\label{Fig:results_5a}
\end{figure}

\clearpage

\subsection{\label{sec:level2}Demagnetizing field \\ in YIG|Ga:YIG Lieb lattice}

The difficulty in designing the magnonic system is not only due to the adjustment of geometrical parameters of the system but also due to the shaping of the internal magnetic field $\textbf{H}_{\rm eff}$. 

The components of the effective magnetic field can be divided into long-range and short-range. The realization of our model is inseparably linked to the long-range dipole interactions through which the coupling between inclusions is possible. This kind of interaction is sensitive to the geometry of the ferromagnetic elements forming the magnonic system.

In Lieb lattice, the nodes of minority sublattice $A$ have four neighbours and the nodes of majority sublattice $B$ have two. As a result, identical inclusions (in terms of their shapes and material parameters) become distinguishable, because of slightly different values of the internal demagnetising field. This has consequences for the formation of a frequency gap between Dirac cones at point $M$ in the dispersion relation obtained for the basic Lieb lattice. In the literature, this phenomenon has been described for the tight-binding model and is called node dimerisation of the lattice \cite{jiang_lieb-like_2019}.

\begin{figure}[!ht]
\includegraphics[width=8cm]{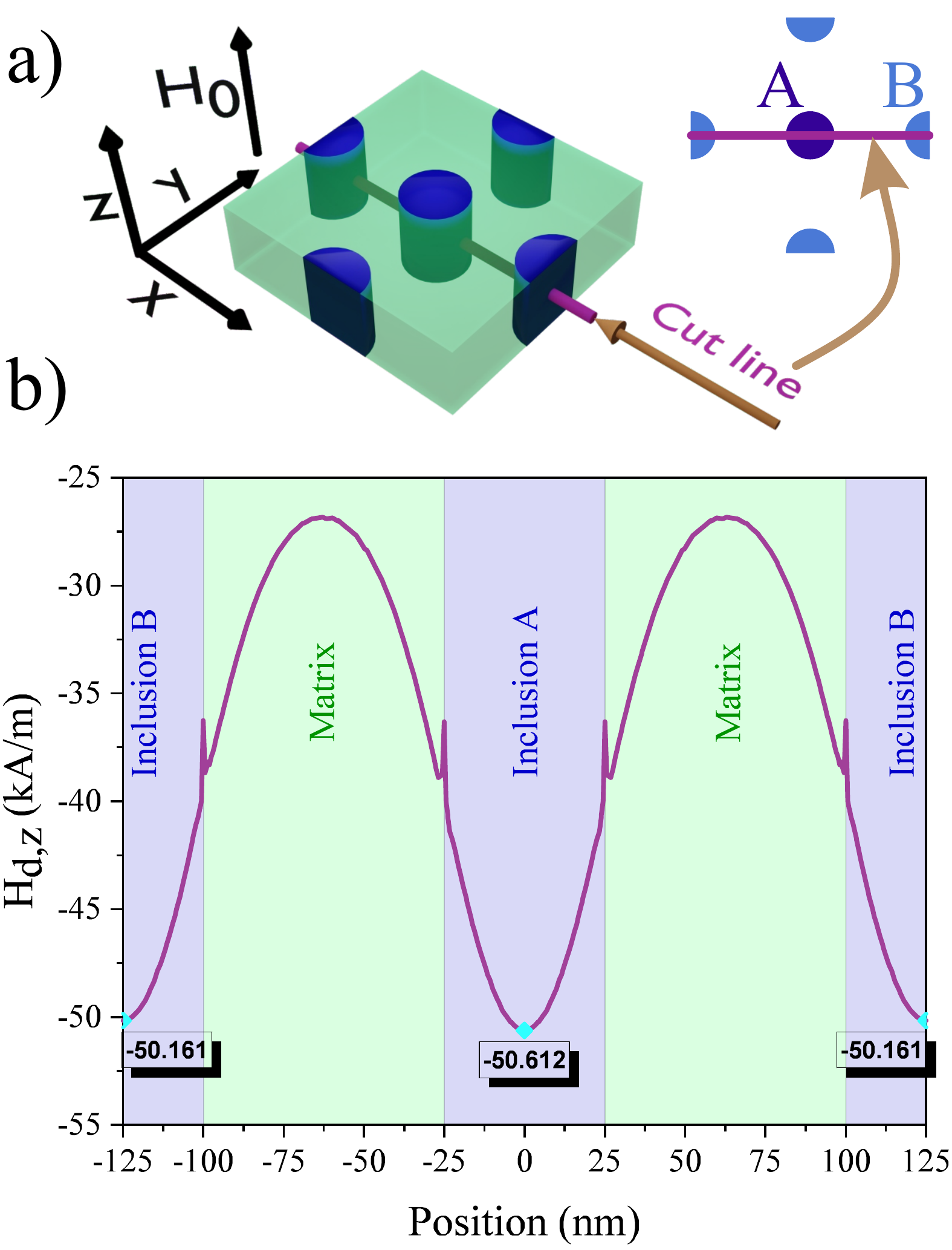}
   \caption{\small{Profile of static demagnetizing field plotted at cut through (a) Lieb lattice unit cell. (b) The $z$-component of the demagnetizing field along the cut line is shown in (a). In the plot, we have marked peaks for the areas of inclusions  $A$ and $B$. Please note the slightly different values of demagnetizing in the centre of $A$ and $B$ inclusion due to different the number neighboring of nodes: four for inclusion $A$, two for inclusion $B$. } }
\label{Fig:results_Hdz}
\end{figure}

In Fig.~\ref{Fig:results_Hdz} we have shown the profile of the $z$-component of the demagnetising field. For each inclusion through which the cut line passes, we have marked the minimum value of the demagnetising field. The slightly lower value of internal filed for inclusions $A$ is responsible for a tiny lowering of the frequency for the mode 
$M_1$ (concentrated in inclusions $A$) respect the degenerated modes $M_2$ and $M_3$ (confined in inclusions $B$).
\clearpage


\end{document}